# Enhancing Domain Wall Speed in Nanowires with Transverse Magnetic Fields

Andrew Kunz and Sarah C. Reiff

Physics Department, Marquette University, Milwaukee, WI 53233 USA

Dynamic micromagnetic simulation studies have been completed to observe the motion of a domain wall in a magnetic nanowire in an effort to increase the field-driven domain wall speed. Previous studies have shown that the wire dimensions place a cap on the maximum speed attainable by a domain wall when driven by a magnetic field placed along the direction of the nanowire. Here we present data showing a significant increase in the maximum speed of a domain wall due to the addition of a magnetic field placed perpendicular to the longitudinal driving field. The results are expressed in terms of the relative alignment of the transverse field direction with respect to the direction of the magnetic moments within the domain wall. In particular, when the transverse field is parallel to the magnetic moments within the domain wall, the velocity of the wall varies linearly with the strength of the transverse field increasing by up to 20%. Further examination of the domain wall structure shows that the length of the domain wall also depends linearly on the strength of the transverse field. We present a simple model to correlate the effects.

75.60.Ch, 75.60.Jk, 75.70.Kw, 75.75.+a

## Introduction

The speed at which a domain wall can travel in a wire has an impact on the viability of many proposed technological applications in sensing, storage and logic operation[1,2]. When a domain wall is driven by a magnetic field parallel to the long axis of the wire, the maximum wall speed is found to be function of, and capped by, the dimensions of the wire[3-5]. In addition, the range of applied driving fields for which the wall motion is pure, fast translation is also limited by the wire geometry. The critical field at which wall velocity is the greatest is called the Walker field, above this field the wall progresses slowly along the wire in a series of periodic steps[6-8].

Previous attempts have been made to increase the maximum domain wall speed and the value of the critical field[5,9]. The addition of edge roughness in a nanowire has been shown to increase the critical field by effectively suppressing the processional coherence of the magnetic moments in the wire[9]. This technique is essentially the same as increasing the magnetic damping which has also been shown to increase the critical field, however the maximum wall speed remains limited by the dimensions of the wire[5].

In the Walker model, the domain wall speed is given by

$$v = \left(\frac{2\pi\gamma_0\Delta}{\alpha}\right) H_{drive} \qquad (1)$$

where the constant $\gamma_0$ is the gyromagnetic ratio and the constant $\alpha$ is the magnetic damping parameter. The only variable parameters in the expression are the domain wall length $\Delta$ (extent along the wires length) and $H_{drive}$, the longitudinal driving field but the velocity only increases with the driving field up to the critical Walker field[3]. In this work we attempt to increase the maximum domain wall speed by applying a magnetic field component transverse to the driving field. The application of the transverse field varies the energy landscape of the domain wall which leads to a change in the domain wall length. This dynamic domain wall length is dependent on the strength and direction of the transverse field.

In Fig. 1 the results of the micromagnetic simulations completed with and without a transverse field are presented. The relative direction of the transverse field to the direction of the magnetic moments in domain wall is the relevant parameter. The cartoon shows anti-parallel (positive transverse field) alignment. When the transverse field is parallel the domain wall moves at an increased speed from the case of no transverse field and when the transverse field is anti-parallel; which is actually slower. Two other effects are observable in Fig. 1 as well. First, the inclusion of the transverse field does not effect the value of the critical driving field. This is somewhat surprising as it might be assumed that the transverse field would inhibit the nucleation of the vortex associated with the wall motion above the critical field. Second, the wall average wall speed is unchanged above the critical field. This is due to the fact that the nucleated vortex reverses the direction of the magnetic moments in the domain wall[5,8]. This means that the transverse field oscillates between parallel and anti-parallel alignment with the moments in the domain wall, effectively leaving the average motion unchanged.

In Fig. 2 we show the relationship between the domain wall velocity and strength of the transverse field for a 15 Oe driving field in a 100 x 5 nm cross-section wire. The change in speed is shown to be in direct correlation with the change in domain wall length.

**Simulation and Analysis**

Micromagnetic simulation is used to observe the motion of the domain wall within the nanowire[10, 11]. The simulated nanowires are permalloy, 5 microns long with a rectangular cross-section with dimensions ranging from 50 – 200 nm in width and 5 - 20 nm thick. A domain wall is artificially placed near one end of the wire and driven to the other side by an external field applied parallel to the length of the nanowire. In this work an additional field is placed on the wire transverse to the driving field. The alignment of the transverse field to the direction of the moments in the domain wall is the important parameter. As shown in Fig. 3 the moments within the domain wall point down in our simulations. The positive direction is defined to be up, therefore a negative transverse field corresponds to parallel alignment and a positive transverse field is anti-parallel. Figs. 1 and 2 show a transverse field aligned parallel leads to an increase in domain wall speed and domain wall length compared to the zero transverse field and anti-parallel alignment cases.

We calculate the average domain wall speed by finding the time it takes for the domain wall to travel across the central 2.5 micron region of the full 5 micron long wire. This allows us to avoid interactions with the ends of the wire, allows us to ignore the initial acceleration of the wall when the field is applied and gives a long enough range of uniform motion at a constant speed.

To find the domain wall length we take a snapshot of the wall as it passes through the center of the wire. We then plot the component of the reduced long-axis magnetization for each magnetic moment along the wire as a function of position as shown in Fig. 3. The domain wall length is the spatial region where the component of the long-axis magnetization is less than 90% of the value in the bulk of the wire. The plot in Fig. 3 is for the bottom row of magnetic moments in the simulated image of the middle of the nanowire. The region between the dashes is considered to be the domain wall for that row. This calculation is completed for each row of magnetic moments and the domain wall length is considered to be the average of each of the rows. It is clear from Fig. 3 that the domain wall length is not constant throughout the width of the wire, but the general shape is consistent across all trials.

## Results

Fig. 2 shows that the domain wall length and the domain wall speed are correlated and for modest transverse fields each depends linearly on the strength of the transverse field. In the following we present a simple model which demonstrates this relationship. The length of the domain wall is determined by minimizing the relevant energies of the problem. The domain wall is a Neel wall and the energy density of the wall can be expressed as a sum of the energy components $\sigma_{wall} = \sigma_{exchange} + \sigma_{demag} + \sigma_{ext}$ where the individual energy densities are approximated by

$$\sigma_{exchange} = \frac{\pi A}{2\Delta}, \qquad (2a)$$

$$\sigma_{demag} = 2\pi M_s^2 \Delta, \qquad (2b)$$

$$\text{and } \sigma_{ext} = -M_s H_{trans} \Delta. \qquad (2c)$$

*A* and *M_s* are the exchange constant and saturation magnetization of the material respectively, $\Delta$ is the domain wall length, and $H_{trans}$ is the transversely applied magnetic field with the negative sign favoring parallel alignment. Carrying out the minimization with respect to the domain wall length generates an expression showing the linear relationship between the length and the applied field

$$\Delta(H_{trans}) = \Delta_0 + \left(\frac{\Delta_0}{4\pi M_s^2}\right) H_{trans}. \quad (3)$$

In this expression, $\Delta_0$ is the domain wall length for zero transverse field which is also calculated by doing the minimization without(2b). Parallel alignment ($H_{trans} > 0$ in this case) of the field and wall moments increases the domain wall length by reducing the wall energy density. Replacing the static $\Delta$ in (1) with the dynamic $\Delta(H_{trans})$ in (3) shows that both the wall length and wall velocity depend linearly on the transverse field. The simulated results of Fig. 2 confirm this behavior. The result (3) is expressed most conveniently in terms of the zero-field wall length $\Delta_0$ which is experimentally measurable and is strongly dependent on the wire dimensions[6]. It is known that wider wires will have larger domain walls which will in turn have larger wall velocities and according to (4) a greater change in velocity for a given change in transverse field[5]. The simulation results shown in Fig.4 exhibit this behavior correctly. It should be noted that for large parallel transverse fields the linear relationship breaks down and the wall speed and length appear to saturate representing a new speed limit for the domain wall.

## Conclusion

Transverse fields aligned with the direction of the magnetic moments within a nanowire significantly increase the speed of a domain wall for a given driving field below the Walker breakdown field.  A simple model shows that it is the change in the domain wall length that effects the change in domain wall speed.  Larger wires, with larger domain walls show an increasing effect.

## Acknowledgements

This work has been supported by a Cottrell College Science Award from Research Corporation.  We would also like to acknowledge the support of E. Dan Dahlberg and discussions with R. Mattheis in this work.

**Figures and Captions**

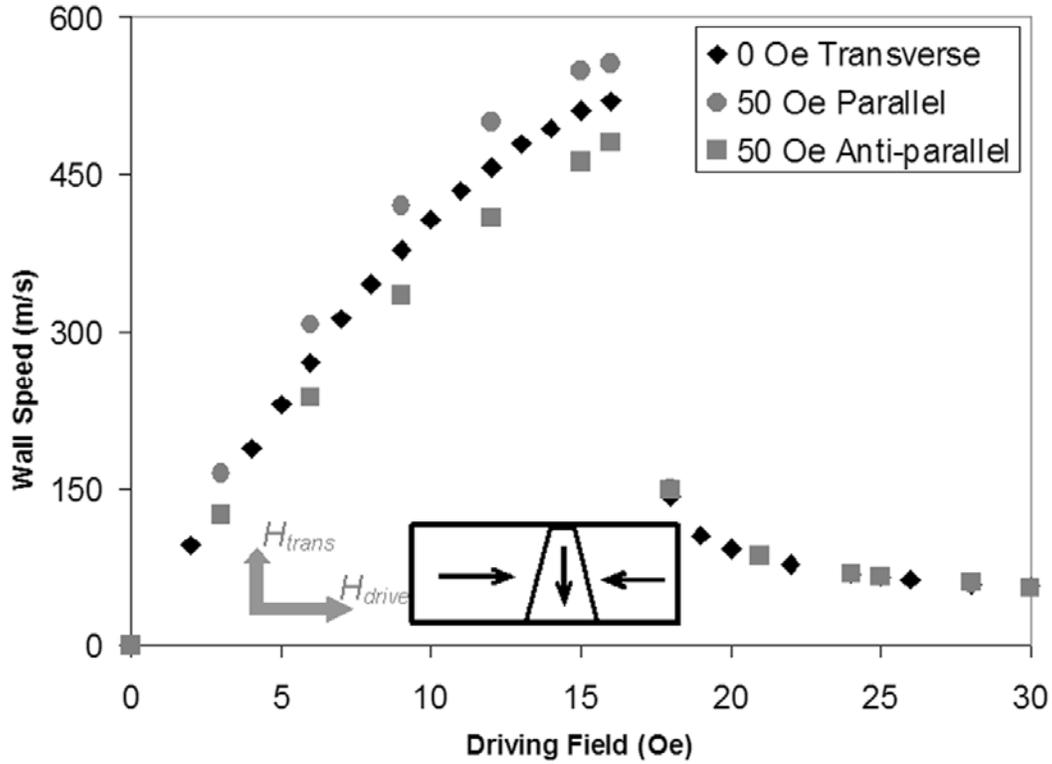

Fig. 1. Average domain wall speed in a 100 x 5 nm cross-section nanowire. The wall speed is shown to increase when a transverse field (-50 Oe) is aligned with the primary direction of the magnetic moments within the domain wall. Anti-parallel alignment results in slower wall speeds. No change is apparent in either the value of the critical field or the average wall speed above the critical field (>16 Oe) when the transverse field is applied.

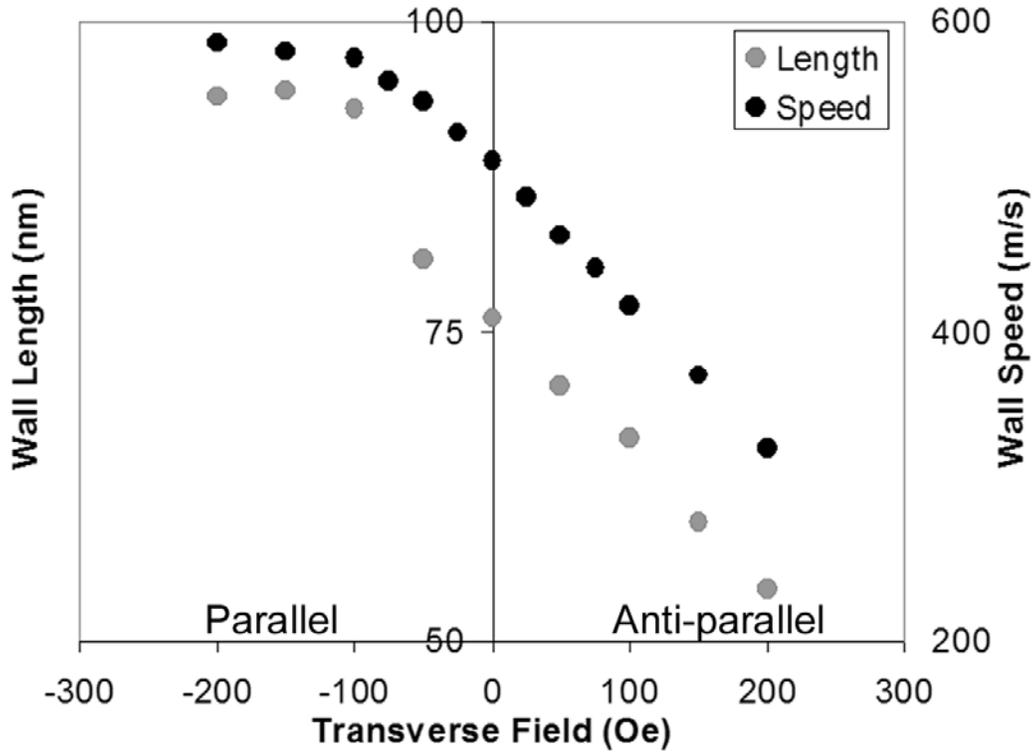

Fig. 2. Domain wall speed and domain length in a 100 x 5 nm cross-section nanowire as a function of transverse field strength and direction for a 15 Oe driving field. The wall speed is correlated to the wall length and both depend linearly on the strength and direction of the transverse field for modest strengths.

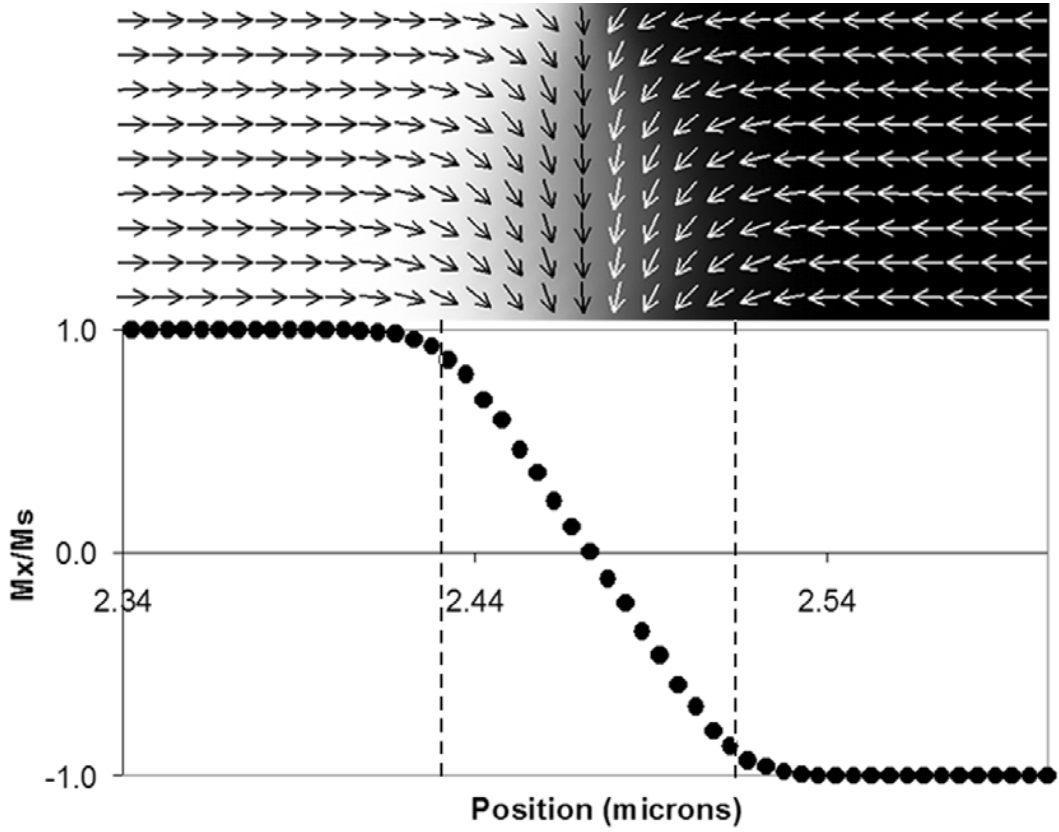

Fig. 3. Domain wall length in a 100 x 5 nm wire. The domain wall is defined to be the region where the reduced long axis magnetization is less than 90% of the value within the bulk of the wire. The plot represents the bottom row of moments in the wire shown at the top.

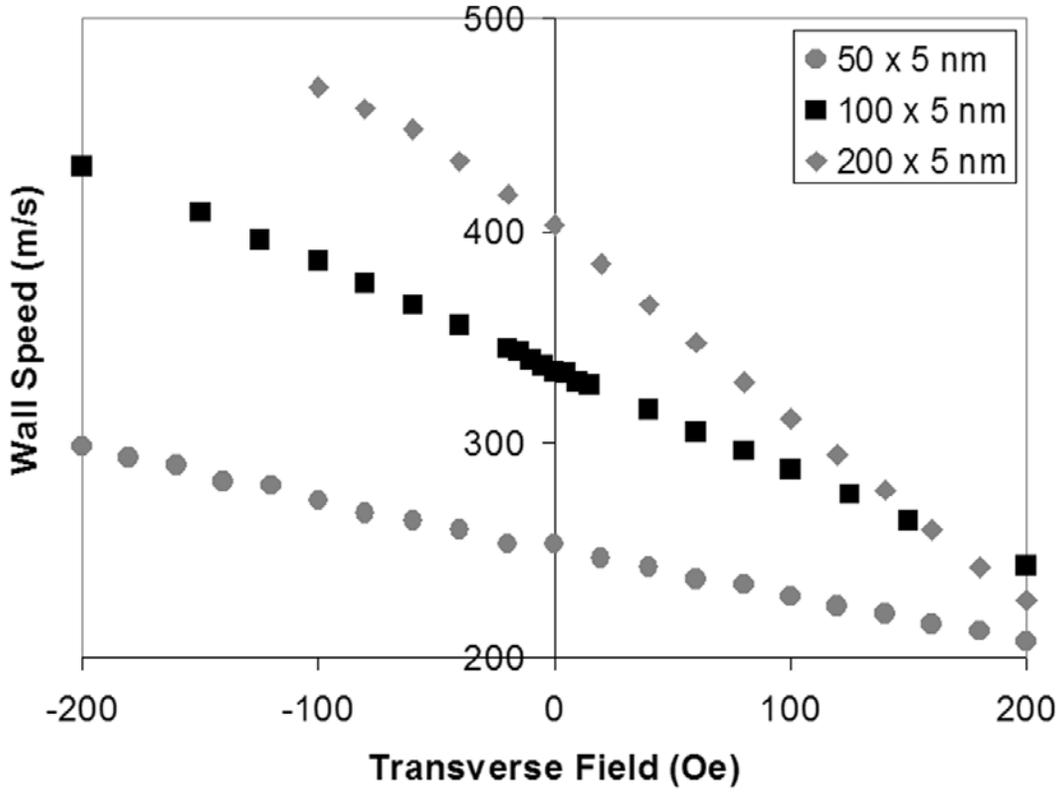

Fig.4 Wall speed vs. transverse field for three different wire dimensions each driven by a 10 Oe field. The zero field speed increases with wire width as does the change in speed with change in transverse field.